\documentclass[aps,prd,floats,floatfix,twocolumn]{revtex4-1}

\usepackage{verbatim}
\usepackage{amsmath}
\usepackage{amsfonts}
\usepackage{amssymb}
\usepackage{mathrsfs}
\usepackage{hyperref}
\usepackage{rotating}
\usepackage{upgreek}
\usepackage{bm}
\usepackage{dcolumn}
\usepackage{epsfig}
\usepackage{graphicx}
\usepackage{graphics}
\usepackage[latin1]{inputenc}
\usepackage{latexsym}
\usepackage[usenames]{color}
\usepackage{yfonts}

\usepackage{graphicx}
\usepackage{xspace} 
\usepackage[usenames]{color}
\usepackage{ulem}
\definecolor {darkgreen}{rgb}{0.2,0.7,0.2}

\newcommand{\be}{\begin{equation}}
\newcommand{\ee}{\end{equation}}
\newcommand{\ben}{\begin{equation*}}
\newcommand{\een}{\end{equation*}}
\newcommand{\bea}{\begin{eqnarray}}
\newcommand{\eea}{\end{eqnarray}}
\newcommand{\ba}{\begin{eqnarray}}
\newcommand{\ea}{\end{eqnarray}}
\newcommand{\bse}{\begin{subequations}}
\newcommand{\ese}{\end{subequations}}
\newcommand{\bal}{\begin{align}}
\newcommand{\eal}{\end{align}}
\newcommand{\nn}{\nonumber}
\newcommand{\bml}{\begin{multline}}
\newcommand{\eml}{\end{multline}}
\newcommand{\bean}{\begin{eqnarray*}}
\newcommand{\eean}{\end{eqnarray*}}

\newcommand{\cd}{\nabla}

\newcommand{\scO}{\mathcal{O}}

\newcommand{\ISCO}{{\mbox{\tiny ISCO}}}

\newcommand{\GW}{{\mbox{\tiny GW}}}
\newcommand{\eff}{{\mbox{\tiny eff}}}

\newcommand{\GR}{{\mbox{\tiny GR}}}
\newcommand{\EH}{{\mbox{\tiny EH}}}

\newcommand{\MAT}{{\mbox{\tiny mat}}}

\newcommand{\CS}{{\mbox{\tiny CS}}}

\newcommand{\starR}{{{}^\ast\!}R}
\newcommand{\pont}{\,\starR\,R}
\newcommand{\met}{\mbox{g}}

\begin{document}

  \title{Non-Spinning Black Holes in Alternative Theories of Gravity}
\author{Nicol\'as Yunes}
  \affiliation{Department of Physics and MIT Kavli Institute, 77
    Massachusetts Avenue, Cambridge, MA 02139}
  \author{Leo C. Stein}
  \affiliation{Department of Physics and MIT Kavli Institute, 77
    Massachusetts Avenue, Cambridge, MA 02139}

\begin{abstract}
We study two large classes of alternative theories, modifying the action through algebraic, quadratic curvature invariants coupled to scalar fields. We find one class that admits solutions that solve the vacuum Einstein equations and another that does not. In the latter, we find a deformation to the Schwarzschild metric that solves the modified field equations in the small coupling approximation. We calculate the event horizon shift, the innermost stable circular orbit shift, and corrections to gravitational waves, mapping them to the parametrized post-Einsteinian framework.
\end{abstract}
\pacs{04.50.Kd, 04.70.-s, 04.80.Cc, 04.30.-w}

\maketitle

\section{Introduction}

Although black holes (BHs) are one of the most
striking predictions of General Relativity (GR), they remain one of
its least tested concepts. Electromagnetic observations have allowed
us to infer their existence, but direct evidence of their
non-linear gravitational structure remains elusive. In the next decade, data from
very long-baseline interferometry~\cite{2009ApJ...695...59D,Fish:2009ak} 
and gravitational wave (GW) 
detectors~\cite{Eardleyprl,Will:2004xi,Berti:2004bd,2005ApJ...618L.115J,Alexander:2007kv,Arun:2009pq,Sopuerta:2009iy,Schutz:2009tz,Yunes:2009bv,Yunes:2009ke,2009PhLB..679..401A,2010GWN.....4....3S,Mishra:2010tp,Yunes:2010yf,Stein:2010pn,2010PhRvD..82l2003T,DelPozzo:2011pg,Molina:2010fb} 
should allow us to image and study BHs in detail. Such
observations will test GR in the dynamical, non-linear or strong-field regime,
precisely where tests are currently lacking.

Testing strong-field gravity features of GR is of utmost importance to
physics and astrophysics as a whole. This is because the particular
form of BH solutions, such as the Schwarzschild and Kerr metrics,
enter many calculations, including accretion disk structure,
gravitational lensing, cosmology and GW
theory. The discovery that these metric solutions do not
accurately represent real BHs could indicate a strong-field departure
from GR with deep implications to fundamental theory.

Such tests require parametrizing deviations from Schwarzschild or
Kerr. One such parameterization at the level of the metric is that of  
{\emph{bumpy BHs}}~\cite{Collins:2004ex,Vigeland:2009pr,Vigeland:2010xe}, 
while another at the level of the GW observable is the {\emph{parameterized post-Einsteinian}}
(ppE) framework~\cite{2009PhRvD..80l2003Y,2010PhRvD..82h2002Y}. In both cases, such parameterizations
are greatly benefited from knowledge of specific non-GR solutions, 
but few, 4D, analytic ones are known that represent regular BHs 
(except perhaps in dynamical Chern-Simons (CS) 
gravity~\cite{2009PhRvD..79h4043Y,Alexander:2009tp} 
and Einstein-Dilaton-Gauss-Bonnet (EDGB) 
gravity~\cite{Kanti:1995vq,Torii:1996yi,Kanti:1997br,Pomazanov:2003wq,Pani:2009wy}).

Most non-GR BH solutions are known through numerical studies. In this
approach, one chooses a particular alternative theory, constructs the
modified field equations and then postulates a metric ansatz with
arbitrary functions. One then derives differential equations for such arbitrary functions 
that are then solved and studied numerically. 
Such an approach was used, for example, to study BHs in
EDGB gravity~\cite{Kanti:1995vq,Torii:1996yi,Kanti:1997br,Pomazanov:2003wq,Pani:2009wy}.

Another approach is to find non-GR BH solutions analytically through
approximation methods. In this scheme, one follows the same route as
in the numerical approach, except that the differential equations for
the arbitrary functions are solved analytically through the aid of approximation 
methods, for example by expanding in (a dimensionless function of) the coupling constants of the
theory. Such a {\emph{small-coupling approximation}}~\cite{Campanelli:1994sj,Cooney:2008wk,2009PhRvD..79h4043Y} treats the
alternative theory as an {\emph{effective and approximate}} model that
allows for small GR deformations. This approach has been used 
to find an analytic, slowly-rotating BH solution in dynamical CS
modified gravity~\cite{2009PhRvD..79h4043Y,Alexander:2009tp}.

But not all BH solutions outside of GR must necessarily be different from standard GR ones. 
In fact, there exists many modified gravity theories where the Kerr metric remains a solution. 
This was the topic studied in~\cite{2008PhRvL.100i1101P}, where it was explicitly shown that
the Kerr metric is also a solution of certain $f(R)$ theories, non-dynamical quadratic gravity 
theories, and certain vector-tensor gravity theories. Based on these fairly generic examples, 
it was then inferred that the astrophysical observational verification of the Kerr metric 
could not distinguish between GR and alternative theories of gravity.

Such an inference, however, is not valid, as it was later explicitly shown in~\cite{2009PhRvD..79h4043Y}. Indeed, there are alternative gravity theories, such as dynamical
CS modified gravity, where the Kerr metric is not a solution. This prompted us to
study what class of modified gravity theories admit Kerr and which do not.
We begin by considering the most general quadratic gravity theory with dynamical
couplings, as this is strongly motivated by low-energy effective string
actions~\cite{1992PhLB..285..199C,Boulware:1985wk,Green:1987mn,Green:1987sp,lrr-2004-5}.
When the couplings are static, we recover the results of~\cite{2008PhRvL.100i1101P},
while when they are dynamic we find that the Kerr metric is not a solution.
In the latter case, we find how the Schwarzschild metric must be modified to satisfy the corrected
field equations. We explicitly compute the shift in the location of the event horizon and innermost stable circular orbit.

Such modifications to the BH nature of the spacetime induce corrections to the waveforms
generated by binary inspirals. We compute such modifications and show that they are of
so-called second post-Newtonian (PN) order, i.e.~they correct the GR result at ${\cal{O}}(v^{4})$
relative to the leading-order Newtonian term, where $v$ is the orbital velocity. We further show
that one can map such corrections to the parameterized post-Einsteinian (ppE)
framework~\cite{Yunes:2009ke}, which proposes a model-independent, waveform family that
interpolates between GR and non-GR waveform predictions. This result
supports the suggestion that the ppE scheme can handle a large class
of modified gravity models.

The remainder of this paper is organized as follows. 
Sec~\ref{sec:quad-grav} defines the set of theories we will investigate and computes the
modified field equations.
Sec.~\ref{sec:non-spin-sol} solves for BH solutions in this class of theories.
Sec.~\ref{sec:prop-of-sol} discusses properties of the solution and Sec.~\ref{sec:impact}
studies the impact that such BH modifications will have on the GW observable.
Sec.~\ref{sec:future-work} concludes by pointing to future possible research directions. 
For the remainder of this paper, we use the following conventions:
latin letters in index lists stand for spacetime indices; parentheses and brackets in index lists stand for symmetrization and antisymmetrization respectively, i.e.~$A_{(ab)} = (A_{ab} + A_{ba})/2$ and $A_{[ab]} = (A_{ab} - A_{ba})/2$; we use geometric units with $G = c = 1$.

\section{Quadratic Gravity}
\label{sec:quad-grav}

Consider the wide class of alternative theories of 
gravity in 4-dimensions defined by modifying the Einstein-Hilbert action through all
possible quadratic, algebraic curvature scalars, multiplied by
constant or non-constant couplings: 
\ba S &\equiv& \int d^4x \sqrt{-g} \left\{ \kappa R + \alpha_{1}
f_{1}(\vartheta) R^{2} + \alpha_{2} f_{2}(\vartheta) R_{ab} R^{ab}
\right. \nonumber \\ &+& \left. \alpha_{3} f_{3}(\vartheta) R_{abcd}
R^{abcd} + \alpha_{4} f_{4}(\vartheta) R_{abcd} {}^{*}R^{abcd}
\right. \nonumber \\ &-& \left. \frac{\beta}{2} \left[\nabla_{a}
\vartheta \nabla^{a} \vartheta + 2 V(\vartheta) \right] +
{\cal{L}}_{\rm mat} \right\}\,,
\label{action} \ea
where $g$ is the determinant of the metric $g_{ab}$, $(R, R_{ab},
R_{abcd}, {}^{*}R_{abcd})$ are the Ricci scalar and tensor, the Riemann
tensor and its dual~\cite{Alexander:2009tp} respectively,
${\cal{L}}_{\rm mat}$ is the Lagrangian density for other matter,
$\vartheta$ is a scalar field, $(\alpha_{i},\beta)$ are coupling
constants and $\kappa = (16 \pi G)^{-1}$. All
other quadratic curvature terms are linearly dependent, e.g.~the Weyl tensor squared.
Theories of this type are motivated from fundamental physics, such 
as in low-energy expansions of string theory~\cite{Boulware:1985wk,Green:1987mn,Green:1987sp,lrr-2004-5}.

Let us distinguish between two different types of theories:
non-dynamical and dynamical. In the former, all the couplings are
constant ($f^\prime_{i}(\vartheta) = 0$) and there is no scalar field ($\beta
= 0$). Varying Eq.~\eqref{action} with respect to the metric and setting $f_{i}(\vartheta) = 1$,
we find the modified field equations
\be G_{ab} + \frac{\alpha_{1}}{\kappa} {\cal{H}}_{ab} +
\frac{\alpha_{2}}{\kappa} {\cal{I}}_{ab} + \frac{\alpha_{3}}{\kappa}
{\cal{J}}_{ab} = \frac{1}{2 \kappa} T_{ab}^{\MAT}\,,
\label{nondyn-FEs} \ee
where $T_{ab}^{\MAT}$ is the stress-energy of matter and
\begin{subequations} 
\ba 
{\cal{H}}_{ab} &\equiv& 2 R_{ab} R -
\frac{1}{2} g_{ab} R^{2} - 2 \nabla_{ab} R + 2 g_{ab} \square R\,,
\label{non-dyn-funcs-a} 
\\
{\cal{I}}_{ab} &\equiv& \square R_{ab} + 2
R_{acbd} R^{cd} - \frac{1}{2} g_{ab} R_{cd} R^{cd} 
\nonumber \\
&+& \frac{1}{2} g_{ab} \square R - \nabla_{ab} R\,,
\ea
\ba
 {\cal{J}}_{ab}
&\equiv& 8 R^{cd} R_{acbd} - 2 g_{ab} R^{cd} R_{cd} + 4 \square R_{ab}
\nonumber \\ &-& 2 R \, R_{ab} + \frac{1}{2} g_{ab} R^{2} - 2 \nabla_{ab}
R\,,
\label{non-dyn-funcs-c} \ea
\end{subequations}
with $\nabla_{a}$, $\nabla_{ab} = \nabla_{a} \nabla_{b}$ and $\square = \nabla_{a} \nabla^{a}$ the
first and second covariant derivatives and the d'Alembertian, and using the
Weyl identity $4C_{a}{}^{cde} C_{bcde}$ $= g_{ab} C_{cdef} 
C^{cdef}$, with $C_{abcd}$ the Weyl tensor.

The dynamical theory is specified through the action in
Eq.~\eqref{action} with $f_{i}(\vartheta)$ some function of the
dynamical scalar field $\vartheta$, with potential $V(\vartheta)$. For
simplicity, we restrict attention here to functions that admit the Taylor expansion
$f_{i}(\vartheta) = f_{i}(0)+ f_{i}^\prime(0) \vartheta + {\cal{O}}(\vartheta^{2})$
about small $\vartheta$, where $f_{i}(0)$ and $f_{i}^\prime(0)$ are constants.
The $\vartheta$-independent terms, proportional to $f_{i}(0)$, 
lead to the non-dynamical theory, and we thus ignore them
henceforth. Let us then concentrate on $f_{i}(\vartheta) = c_{i} \vartheta$, where we reabsorb the 
constants $c_{i} = f_{i}^\prime$ into $\alpha_{i}$, such that 
$\alpha_{i} f_{i}(\vartheta) \to  \alpha_{i} \vartheta$. The field equations are then
\ba G_{ab} &+& \frac{\alpha_{1}}{\kappa}
{\cal{H}}_{ab}^{(\vartheta)} + 
\frac{\alpha_{2}}{\kappa} {\cal{I}}_{ab}^{(\vartheta)} \\ \nonumber
&+& \frac{\alpha_{3}}{\kappa} {\cal{J}}_{ab}^{(\vartheta)} +
\frac{\alpha_{4}}{\kappa} {\cal{K}}_{ab}^{(\vartheta)} 
= \frac{1}{2\kappa} \left( T_{ab}^{\MAT} + T_{ab}^{(\vartheta)} \right)\,,
 \label{dyn-FEs} \ea
where $T_{ab}^{(\vartheta)} = \frac{\beta}{2}\left[\nabla_{a} \vartheta \nabla_{b}
\vartheta - \frac{1}{2}g_{ab} \left(\nabla_{c} \vartheta \nabla^{c} \vartheta - 2
V(\vartheta) \right) \right]$ is the scalar field stress-energy tensor
and
\begin{subequations}
\ba
{\cal{H}}_{ab}^{(\vartheta)} &\equiv& -4
v_{(a} \nabla_{b)} R - 2 R \nabla_{(a} v_{b)} + g_{ab} \left(2
R\nabla^{c} v_{c} + 4 v^{c} \nabla_{c}R \right) \nn \\
&+& \vartheta \left[ 2 R_{ab} R - 2\nabla_{ab} R - \frac{1}{2}
  \met_{ab} \left( R^2 - 4\square R \right) \right] \,,
\\
{\cal{I}}_{ab}^{(\vartheta)} &\equiv& -v_{(a} \cd_{b)} R - 2v^c \left(
  \cd_{(a} R_{b)c} - \cd_c R_{ab} \right) + R_{ab} \cd_c v^c  
  \nonumber \\
&-& 2 R_{c(a}\cd^c v_{b)} + \met_{ab} \left( v^c
  \cd_c R + R^{cd} \cd_c v_d \right) \nn\\
&+& \vartheta \left[ 2 R^{cd} R_{acbd} - \cd_{ab} R + \square R_{ab}
\right. \nonumber \\
&+& \left. \textstyle{\frac{1}{2}}\met_{ab} \left(  \square R - R_{cd} R^{cd}
  \right) \right] \,,
  \\
{\cal{J}}_{ab}^{(\vartheta)}
&\equiv& - 8 v^c \left( \nabla_{(a} R_{b)c} - \nabla_{c} R_{ab}\right) + 4
R_{acbd} \nabla^{c} v^{d} 
\nn \\
&-& \vartheta \left[ 2 \left( R_{ab} R - 4R^{cd} R_{acbd} + \nabla_{ab}
  R - 2 \square R_{ab} \right) \right. \nonumber\\
&-& \left. \textstyle{\frac{1}{2}} \met_{ab} \left( R^2 - 4 R_{cd}
    R^{cd}\right) \right] \,, \nn\\
{\cal{K}}_{ab}^{(\vartheta)}
&\equiv& 4 v^{c} \epsilon_{c}{}^{d}{}_{e(a} \nabla^{e} R_{b)d} + 4
\nabla_{d}v_{c} {}^{*}R_{(a}{}^c{}_{b)}{}^d\,,
\ea
\end{subequations}
with $v_{a} \equiv \nabla_{a} \vartheta$ and $\epsilon^{abcd}$ the
Levi-Civita tensor. Notice that $\alpha_{4} {\cal{K}}_{ab} =
\alpha_{\CS} C_{ab}$, where $\alpha_{\CS}$ and $C_{ab}$ are the 
CS coupling constant and the CS C-tensor~\cite{Alexander:2009tp}. 
The dynamical quadratic theory includes dynamical
CS gravity as a special case. Variation of the action with respect
to $\vartheta$ yields the scalar field equation of motion
\ba \beta \square \vartheta - \beta \frac{dV}{d\vartheta} &=& -
\alpha_{1} R^{2} - \alpha_{2} R_{ab} R^{ab} \nonumber \\ &-&
\alpha_{3} R_{abcd} R^{abcd} - \alpha_{4} R_{abcd} {}^{*}R^{abcd}\,.
\label{EOM} \ea

Both the non-dynamical and dynamical theories arise from a
diffeomorphism invariant action, and thus, they lead to field equations
that are covariantly conserved, i.e.~the covariant divergence of
Eq.~\eqref{nondyn-FEs} identically vanishes, while that of
Eq.~\eqref{dyn-FEs} vanishes upon imposition of Eq.~\eqref{EOM},
unlike in non-dynamical CS gravity~\cite{Alexander:2009tp}.

\section{Non-Spinning Black Hole Solution}
\label{sec:non-spin-sol}

\subsection{Non-dynamical Theories}

The modified field equations of the non-dynamical theory have the interesting property
that metrics for which the Ricci tensor vanishes are automatically
solutions.  One can see that if $R_{ab} = 0$,
then Eqs.~\eqref{non-dyn-funcs-a}-\eqref{non-dyn-funcs-c} vanish
exactly, thus satisfying the modified field equations in
Eq.~\eqref{nondyn-FEs}. This generalizes the result
in~\cite{2008PhRvL.100i1101P}, as we here considered a more general action.

The reason for this simplification is the Gauss-Bonnet and Pontryagin
identities. The integral of the Gauss-Bonnet term ${\cal{G}} \equiv R^{2} - 4
R_{ab} R^{ab} + R_{abcd} R^{abcd}$ is proportional to the Euler
characteristic ${\cal{E}}$, while that of the Pontryagin density
$R_{abcd} {}^{*}R^{abcd}$ is proportional to the Chern number
${\cal{C}}$. Thus, the $R_{abcd} R^{abcd}$ and the $R_{abcd}
{}^{*}R^{abcd}$ terms can be removed from the action in
Eq.~\eqref{action} in favor of ${\cal{E}}$ and ${\cal{C}}$. Since the
variation of these constants vanishes identically, the field equations
can be rewritten to depend only on the Ricci tensor and its trace.

This feature has a natural generalization for a wider class of alternative
theories of gravity. If an action for an alternative theory contains the Riemann 
tensor or its dual \emph{only} in a form that can be rewritten in
terms of topological invariants (with no dynamical
couplings), then the field equations will be free of Riemann, and thus,
all vacuum GR solutions will also be solutions of such modified theories.
Therefore, any action built from powers of the Ricci scalar or products of the Ricci tensor,
possibly coupled to dynamical fields, and with Riemann tensors
entering only as above, admits all vacuum GR solutions.

These results have important consequences for attempts to test GR in the strong field.  
Electromagnetic GR tests that aim at probing the Kerr nature of BHs would
be insensitive to such modified theories. On the other hand, observations that probe 
the dynamics of the background, such as GW observations~\cite{Will:2004xi,Berti:2004bd,2005ApJ...618L.115J,Alexander:2007kv,Arun:2009pq,Sopuerta:2009iy,Schutz:2009tz,Yunes:2009bv,Yunes:2009ke,2009PhLB..679..401A,2010GWN.....4....3S,Mishra:2010tp,Yunes:2010yf,Stein:2010pn,2010PhRvD..82l2003T,DelPozzo:2011pg,Molina:2010fb}, 
would be able to constrain them.

\subsection{Dynamical Theories}

The modified field equations in the dynamical theory, however, are not
as simple, as clearly they are not satisfied when $R_{ab} = 0$. This
is because ${\cal{J}}_{ab}^{(\vartheta)}$ depends on $\nabla^{c} v^{d}
R_{abcd}$ and ${\cal{K}}_{ab}^{(\vartheta)}$ depends on $\nabla_{d}
v_{c}\, {}^{*}R_{(a}{}^c{}_{b)}{}^d$. Let us search for small deformations away
from the GR Schwarzschild metric that preserve stationarity and
spherical symmetry. The only relevant term here then is ${\cal{J}}_{ab}^{(\vartheta)}$,
as ${\cal{K}}_{ab}^{(\vartheta)}$  vanishes in spherical symmetry, as already 
analyzed in~\cite{2009PhRvD..79h4043Y}.

We thus pose the ansatz
\be ds^{2} = - f_0 \left[1 + \epsilon h_0(r) \right] dt^{2} + f_0^{-1}
\left[1 + \epsilon k_0(r) \right] dr^{2} + r^{2} d\Omega^{2}\,, \ee
and $\vartheta = \bar{\vartheta} + \epsilon \tilde{\vartheta}$, where
$f_0 \equiv 1 - 2 M_{0}/r$, with $M_{0}$ the ``bare'' or GR BH mass and $(t,r,\theta,\phi)$
are Schwarzschild coordinates, while $d\Omega^{2}$ is the line element on
the 2-sphere. The free functions $(h_0,k_0)$ are small deformations from
the Schwarzschild metric, controlled by a function of the coupling
constants $(\alpha_{i},\beta)$ that we define below; $\epsilon$ is a
book-keeping parameter.


Before we solve the field equations, let us discuss the scalar field
potential $V(\vartheta)$. There are two distinct choices for this
potential: a flat ($V'(\vartheta) = 0$) or non-flat ($V'(\vartheta)
\neq 0$) potential. For the non-flat case, the potential must be
bounded from below for the theory to be globally stable, and thus it
will contain one or more minima. The scalar field would tend towards
the minimum of the potential, where the latter could be expanded as
a quadratic function about the minimum (assumed here to be at zero): 
$V \approx \frac{1}{2}m_\vartheta^2 \vartheta^2$. One
might treat the flat potential as the limit $m_\vartheta \to 0$ of the above
non-flat potential, but this limit is not continuous at the point $m_\vartheta=0$. 
The massive case must thus be treated generically and it turns out to be sufficiently 
complicated that we restrict our attention only to the massless (flat) case\footnote{It is worth noting,
however, that the potential must respect the symmetries inherited from the fundamental
theory that the effective action derives from. A large class of such theories, such as heterotic
string theory in the low-energy limit, is shift symmetric, which then forbids the appearance of 
mass terms.}.

With this ansatz, we can solve the modified field equations and the
scalar field's equation of motion order by order in
$\epsilon$. Through the small-coupling approximation, we treat $\alpha = {\cal{O}}(\epsilon)$
and $\beta = \cal{O}(\epsilon)$. To zeroth-order in $\epsilon$, the
field equations are automatically satisfied because the Schwarzschild
metric has vanishing Ricci tensor. To this order,
the scalar field equation can be solved to find
\be \bar{\vartheta} = \frac{\alpha_{3}}{\beta} \frac{2}{M_{0} r} \left(1 +
\frac{M_{0}}{r} + \frac{4}{3} \frac{M_{0}^{2}}{r^{2}} \right)\,.  
\label{SF-sol}
\ee
This is the same solution found in~\cite{1992PhLB..285..199C} 
for dilaton hair sourced in EDGB gravity. The scalar field
depends only on $\alpha_{3}$, since the term proportional to
$\alpha_{4}$ vanishes identically in a spherically symmetric
background. 

We can use this scalar field solution to solve the
modified field equations to $\scO(\epsilon)$. Requiring that the metric
be asymptotically flat and regular at $r = 2 \, M_{0}$, we find the unique solution
$h_0 \equiv {\cal{F}} (1 + \tilde{h}_0)$ and $k_0 \equiv -{\cal{F}} (1 + \tilde{k}_0)$, where
${\cal{F}} \equiv -(49/40) \; \zeta \; (M_{0}/r)$ and
\ba 
\label{sols-h-m}
\tilde{h}_0 &=& \frac{2 M_{0}}{r} + \frac{548}{147} \frac{M_{0}^{2}}{r^{2}} + \frac{8}{21} \frac{M_{0}^{3}}{r^{3}} -
\frac{416}{147} \frac{M_{0}^{4}}{r^{4}} - \frac{1600}{147} \frac{M_{0}^{5}}{r^{5}}\,, 
\\ \nonumber 
\tilde{k}_0 &=& \frac{58}{49} \frac{M_{0}}{r} + \frac{76}{49} \frac{M_{0}^{2}}{r^{2}}
- \frac{232}{21} \frac{M_{0}^{3}}{r^{3}}  - \frac{3488}{147} \frac{M_{0}^{4}}{r^{4}} -
\frac{7360}{147} \frac{M_{0}^{5}}{r^{5}}\,,  
\ea
and where we have defined the dimensionless coupling function $\zeta
\equiv \alpha_{3}^{2}/(\beta \kappa M_{0}^{4}) = {\cal{O}}(\epsilon)$. This solution 
is the same as that found in EDGB gravity~\cite{Mignemi:1992pm}. Our analysis
shows that such a solution is the most general for all dynamical,
algebraic, quadratic gravity theories, in spherical symmetry.

The demand that the metric deformation be regular everywhere outside the horizon
has led to a term that changes the Schwarzschild BH mass, i.e.~there is a correction
to $g_{tt}$ and $g_{rr}$ that decays as $1/r$ at spatial infinity. We can then define
the physical mass $M \equiv M_{0} [1 + (49/80) \zeta]$, such that the only modified 
metric components become $g_{tt} = -f (1+h)$ and $g_{rr} = f^{-1}
(1+k)$ where $h = \zeta/(3 f) (M/r)^{3} \tilde{h}$ and $k = - (\zeta/f) (M/r)^{2}
\tilde{k}$, and
\ba
\tilde{h} &=& 1 + \frac{26 {M}}{r}
+ \frac{66}{5} \frac{{M}^{2}}{r^{2}} + \frac{96}{5} \frac{{M}^{3}}{r^{3}} 
- \frac{80 {M}^{4}}{r^{4}}\,,
\\
\tilde{k} &=& 1 + \frac{{M}}{r} + \frac{52}{3} \frac{{M}^{2}}{r^{2}}
+ \frac{2 {M}^{3}}{r^{3}} + \frac{16}{5} \frac{{M}^{4}}{r^{4}}
- \frac{368}{3} \frac{{M}^{5}}{r^{5}}\,, \quad
\ea
and where $f \equiv 1 - 2 M/r$.
Physical observables are related on the renormalized mass, not the bare
mass. This renormalization was not performed
by~\cite{Mignemi:1992pm}.

In fact, one need not fix the single constant of integration which
appears in finding this solution. Any value of the integration
constant, after renormalization, is absorbed into the renormalized
mass. Rather than a family of spacetimes, there is a unique spacetime
after renormalization.

The sign of the coupling constant can be determined by computing the energy
carried by the scalar field in Eq.~\eqref{SF-sol}. The energy is
$E_{(\vartheta)} \equiv \int_{\Sigma} T_{ab}^{(\vartheta)} t^{a} t^{b} \gamma^{1/2} d^{3}x$,
where $\Sigma$ is a $t = \rm{const.}$ hypersurface outside of the
horizon (so that it is spacelike everywhere), $t^{a} = (\partial/\partial t)^{a}$ and $\gamma$
is the determinant of the metric intrinsic to $\Sigma$. We find that 
$E_{(\vartheta)} = (9/7) {\zeta} \kappa \pi {M}$. For stability reasons, we require that
$E_{(\vartheta)} \geq 0$, which then implies ${\zeta} \geq 0$ and $\alpha_{3}^{2}/\beta \geq 0$.

Although we here considered non-spinning BHs, our analysis can be generalized
to spinning ones, by separating the theory and its solutions into parity-even and parity-odd 
sectors. A parity transformation consists of the reflection $x^{i} \to -x^{i}$, which for 
a spinning BH metric implies $a\to -a$, where $|S^{i}| = M |a|$ is the magnitude of the spin 
angular momentum. Expanding the spinning BH solution as a power series
in $a/M$, we see that the Kretschmann scalar $R_{abcd}R^{abcd}$ has only even 
powers of $a/M$ (even parity sector), while the Pontryagin density $\pont$ has only odd 
powers of $a/M$ (odd parity sector). These quantities source the $\vartheta$ equation of 
motion, therefore driving even and odd metric perturbations respectively.  
The solution found here is of even parity and corresponds to the $\scO(a^0)$ part of the
metric expansion for a slowly-spinning BH in dynamical quadratic gravity. 
The next order, $\scO(a^1)$, is parity odd and is
sourced only by the Pontryagin density, since $R^2, R_{ab}R^{ab}$, and
$R_{abcd}R^{abcd}$ are all even under parity. The solution sourced by
just the Pontryagin density is identical to that in dynamical
Chern-Simons gravity (all $\alpha_{i} = 0$ except for $\alpha_4$) and
was found in~\cite{2009PhRvD..79h4043Y}. From the parity arguments presented
here, we see that the exact same modification arises at $\scO(a^1)$ in the more
general dynamical quadratic gravity considered here.
Therefore, to $\scO(a^1)$, the
modification in dynamical quadratic gravity is simply the linear
combination of the $\scO(a^0)$ solution found
here and the $\scO(a^1)$ solution found in~\cite{2009PhRvD..79h4043Y}.

\section{Properties of the Solution}
\label{sec:prop-of-sol}

The solution found is spherically symmetric, stationary, 
asymptotically flat, and regular everywhere except at $r=0$. 
It represents a non-spinning BH with a real singularity at the origin, 
as evidenced by calculating the Kretschmann scalar expanded to 
${\cal{O}}(\zeta)$: $K \equiv R_{abcd} R^{abcd} = \bar{K} - 32 \zeta M^{3}/r^{7} \tilde{K}$, where
$\bar{K} = 48 {M}^{2}/r^{6}$ and
\ba
\tilde{K} &=&1 + \frac{{M}}{2 r} + \frac{72 {M}^{2}}{r^{2}} 
+ \frac{7 {M}^{3}}{r^{3}} + \frac{64}{5} \frac{{M}^{4}}{r^{4}} 
- \frac{840 {M}^{5}}{r^{5}}\,. \quad
\ea
The location of the event horizon, i.e.~the surface of infinite redshift, can be computed by solving
$g_{tt} = 0$ to find $r_{\EH}/{M} = 2  - (49/40) \zeta$. 
The metric remains Lorentzian (i.e.~${\rm sgn}(g)<0$)
everywhere outside $r_{\EH}$ provided $\zeta$ is sufficiently small 
(specifically, $0<\zeta < (120/361)$).

One can also study point-particle motion in this background. Neglecting internal structure and spins, 
test-particle motion remains geodesic~\cite{Gralla:2008fg} and the equation of motion reduces to 
$\dot{r}^{2}/2 = V_{\eff}^{\GR} + \delta V_{\eff}$, where the overhead dot stands for differentiation 
with respect to proper time and 
\be
V_{\eff}^{\GR} = \frac{E^{2}}{2} - \frac{L^{2}}{2 r^{2}} f - \frac{f}{2}\,,
\qquad
\delta V_{\eff} = -\frac{1}{2} E^{2} {h} - \frac{1}{2} V_{\eff}^{\GR} k\,,
\ee
where $(E,L)$ are the conserved quantities induced by the timelike and azimuthal Killing vectors,  i.e.~the particle's energy and angular momentum per unit mass.

One can solve for the energy and angular momentum for circular orbits~\cite{Wald:1984cw} through the 
conditions $\dot{r} = 0$ and $V_\eff^\prime = 0$ to find $E = E_{\GR} + \delta E$ and 
$L = L_{\GR} + \delta L$, where $E_{\GR} = f (1 - 3 {M}/r)^{-1/2}$, 
$L_{\GR} = ({M} r)^{1/2} E_{\GR}/f$ and
\ba
\label{deltaE}
\delta E &=& - \frac{\zeta}{12} \frac{{M}^{3}}{r^{3}} \left(1 - \frac{3 {M}}{r} \right)^{-3/2} 
\left(1 + \frac{54 {M}}{r} + \frac{198}{5} \frac{{M}^{2}}{r^{2}} 
\right. \nonumber \\
&&+ \left. \frac{252}{5} \frac{{M}^{3}}{r^{3}} - \frac{2384}{5} \frac{{M}^{4}}{r^{4}} + \frac{480 {M}^{5}}{r^{5}}\right)\,,
\\
\delta L &=& - \frac{\zeta {M}}{4} \frac{{M}^{3/2}}{r^{3/2}} \left(1 - \frac{3 {M}}{r} \right)^{-3/2}
\left(1 + \frac{100}{3} \frac{{M}}{r}
\right. \nonumber \\
&&- \left.
 \frac{30 {M}^{2}}{r^{2}} + \frac{16}{5} \frac{{M}^{3}}{r^{3}} - \frac{752}{3} \frac{{M}^{4}}{r^{4}} + \frac{320 {M}^{5}}{r^{5}} \right)\,.
\ea
From this expression, we can find the modified Kepler law by expanding $\omega \equiv L/r^{2}$ 
in the far field limit: 
\be
\omega^{2} = \omega_{\GR}^{2} \left[1 - \frac{\zeta}{2} \left(\frac{M}{r} \right)^{2} \right]\,
\ee 
where $\omega_{\GR}^{2} = {M}/r^{3} [1 + {\cal{O}}({M}/r)]$. If in addition to the above circular orbit 
conditions one evaluates the marginal stability condition $V_\eff^{\prime\prime} = 0$, one finds that 
the shift in the ISCO location is 
\be
\frac{r_{\ISCO}}{M} = 6 - \frac{16297}{9720} \zeta\,.
\ee

\section{Impact on Binary Inspiral GWs}
\label{sec:impact}

As evidenced above, such a modified theory
will introduce corrections to the binding energy of binary systems. Consider a binary with
component masses $m_{1,2}$ and total mass $m = m_{1} + m_{2}$. The binding energy, 
to leading ${\cal{O}}(m/r,\zeta)$, can be obtained from $E_{\GR}$ and $\delta E$ in Eq.~\eqref{deltaE}
by the transformation $m_{1} m_{2} \to m^{2} \eta$ and expanding in $M/r \ll 1$. 
This trick works to leading order in $\zeta$ and in $m/r$ only and it leads to
\be
E_{b}(r) = - \frac{m^{2} \eta}{2 r} \left[1+ \frac{\zeta}{6} \left(\frac{m}{r} \right)^2\right]\,.
\ee
Using the modified Keplerian relation of the previous Section, this becomes 
\be
E_{b}(F) = - \frac{1}{2} \left(2 \pi m F \right)^{2/3} - \frac{1}{6} m \eta \zeta \left(2 \pi m F\right)^{2}\,, 
\ee
to leading  ${\cal{O}}(mF,\zeta)$, where $F$ is the orbital frequency and $\eta = m_{1} m_{2}/m^{2}$
is the symmetric mass ratio. Such a modification to the binding energy will introduce corrections to 
the binary's orbital phase evolution at leading, Newtonian order. 

A calculation of the phase and amplitude waveform correction that accounts only for the 
leading-order binding energy modification is incomplete. 
First, higher ${\cal{O}}(m/r)$ terms in $E_{b}$ are necessary for detailed GW tests. 
These terms, however, are not necessary to find the leading-order, functional form of the 
waveform correction; this is all one needs to map these modifications to the ppE scheme.   

To be consistent, we must also consider the energy flux carried by the
scalar field. This program involves solving for the perturbation on
top of the background solution given in Eq.~\eqref{SF-sol}. The
solution can be found using post-Newtonian integration techniques and
is in preparation~\cite{yunes-yagi-prep}. The modification to radiation reaction
due to the scalar field is subdominant (of much higher post-Newtonian
order) compared to the modification to the binding energy calculated here, 
as will be shown in a forthcoming paper~\cite{yunes-yagi-prep}.

Let us now compute the orbital phase correction due to modifications to the binding energy. 
The orbital phase for a binary in a circular orbit is simply  
\be
\phi(F) = \int^{F} \, (E') \, (\dot{E})^{-1}  \omega \, d \omega\,,
\ee
where $\omega=2 \pi F$ is the orbital angular frequency, 
$E' \equiv dE/d\omega$ and $\dot{E}= -(32/5) \eta^{2} m^{2} r^{4} \omega^{6}$
is the loss of binding energy due to radiation.
This expression for $\dot{E}$ is the GR quadrupole form, which was
shown~\cite{Stein:2010pn} to be valid in the small-coupling limit in
asymptotically flat spacetimes when the action is of the form we use.
Neglecting $\dot{E}^{(\vartheta)}$ and to leading ${\cal{O}}(m \omega,\zeta)$,
the orbital phase 
\be
\phi = \phi_{\GR} \left[1 + \frac{25}{3} \zeta \; (2 \pi m F)^{4/3} \right]\,,
\ee
where the GR phase is $\phi_{\GR} = -1/(32 \eta) (2 \pi m F)^{-5/3}$.
The leading order correction is of so-called $2$PN order, as it scales with
$(m \; F)^{4/3}$ (down by $1/c^{4}$) relative to the leading-order GR result. 
 
Similarly, we can compute the correction to the frequency-domain GW phase in the SPA,
by assuming that its rate of change is much more rapid than the GW amplitude's.
This phase is (see e.g.~\cite{Yunes:2009yz})
\be
\Psi_{\GW} = 2 \phi(t_{0}) - 2 \pi f t_{0}\,,
\ee
where $t_{0}$ satisfies the stationary phase condition $F(t_{0}) = f/2$, with $f$ the GW frequency. 
Neglecting $\dot{E}^{(\vartheta)}$ and to leading ${\cal{O}}(m\omega,\zeta)$, we find that
\be
\Psi_{\GW} = \Psi_{\GW}^{\GR} \left[1 + \frac{50}{3} \zeta \eta^{-4/5} u^{4/3} \right]\,,
\ee 
where $u \equiv \pi {\cal{M}} f$ is the reduced frequency and
${\cal{M}} = \eta^{3/5} m$ is the chirp mass. Similarly, the Fourier-domain amplitude scales
as $|\tilde{h}| \propto \dot{F}(t_{0})^{-1/2}$, which then leads to
\be
|\tilde{h}| = |\tilde{h}|_{\GR} \left[1 + \frac{5}{6} \zeta u^{4/3} \eta^{-4/5} \right]\,,
\ee 
where $|\tilde{h}_{\GR}|$ is the GW amplitude in GR. In principle, there could
be additional corrections to $|h|$ from modifications to the first order equations of motion,
but~\cite{Stein:2010pn} has shown that these vanish in the small coupling approximation.

The modifications introduced to the inspiral waveforms can be mapped to parametrized waveform
models that facilitate GR tests. In the ppE framework~\cite{2009PhRvD..80l2003Y},
the simplest parameterization is
\be
\tilde{h} = |\tilde{h}|_{\GR} (1 + \alpha \eta^{c} u^{a}) \exp[i \Psi_\GW^{\GR} (1 + \beta \eta^{d} u^{b})]\,, 
\ee
where $(\alpha,a,\beta,b,c,d)$ are ppE parameters. Our results clearly
map to this parameterization with $\alpha = (5/6) \zeta$, $\beta =
(50/3) \zeta$, $a = 4/3 = b$ and $c = -4/5 = d$.  Since the radiation
carried by the scalar field is of higher post-Newtonian order,
including it will not change these ppE parameters.  Future GW constraint
on these parameters could be translated into a bound on the class of
alternative theories considered here.

Preliminary studies suggest that GW detectors, such as LIGO, could place interesting constraints
on the parameter $\beta$. Given a signal-to-noise ratio of 20 for a comparable mass binary inspiral signal, 
one might be able to constrain $\beta \lesssim 10^{-1}$ when $b = 4/3$~\cite{ppEImplement}. 
This bound would translate to a $\zeta$-constraint of $\zeta \lesssim 10^{-2}$, which should be compared to the
current double binary pulsar constraint $\zeta \lesssim 10^{7}$~\cite{2010PhRvD..82h2002Y}. 
We then see that GWs could place much stronger constraints on non-linear strong field deviations from GR 
relative to current binary pulsar bounds. 

\section{Future Work}
\label{sec:future-work}

The study presented here shows that there is a wide class of modified gravity theories
where Schwarzschild and Kerr are not solutions, yet their waveform modifications can be
mapped to the ppE scheme. This study could be extended by investigating higher-order
in $v$, PN corrections to the waveform modifications. Such a calculation would require one
to solve for the two-body metric in this specific class of theories. Although this can in principle
be done within the PN scheme, in practice the calculation will be analytically quite difficult,
due to the non-linear terms introduced by the modified theory.

Another possible extension is to investigate the effect of different potential terms to the 
results presented here. For example, one could postulate a cosine potential and see how this
modifies the solutions found. Such cosine potentials arise naturally due to non-linear
interactions in effective string actions. The inclusion of such a potential will probably render
the problem non-analytic, forcing us to solve the equations of motion for the scalar field 
numerically. 

One other avenue of future research is to find analytic, closed form solutions for
BHs rotating arbitrarily fast in dynamical quadratic gravity. 
The analysis presented here applies only to non-rotating BHs, and we have discussed how 
it would be modified when considering slowly rotating BHs. 
Exact, closed form solutions for rapidly rotating BHs, however, remain
elusive. One might have to integrate the equations numerically to find such solutions. One
possible line of attack is to evolve the field equations in a $3+1$ decomposition, starting with
a dense and rotating scalar field configuration. Upon evolution, this scalar field will collapse 
into a rapidly rotating BH, yielding a numerical representation of the solution one seeks. 

\acknowledgments

We thank Emanuele Berti, Vitor Cardoso, Stanley Deser, Tim Johannsen, Paolo Pani and
Frans Pretorius for useful comments. NY acknowledges support from NASA through the 
Einstein Postdoctoral Fellowship PF9-00063 and PF0-110080 issued by the Chandra X-ray 
Observatory, which is operated by the SAO for and on behalf of NASA under contract 
NAS8-03060. LCS acknowledges support from NSF Grant PHY-0449884 and 
the Solomon Buchsbaum Fund at MIT. 

\bibliography{master,phyjabb}
\end{document}